\documentclass[flushrt,preprint]{aastex}
\usepackage{amssymb}
\begin{document}
\title{What Perytons Aren't, and Might Be}
\shorttitle{Perytons}
\shortauthors{Katz}
\author{J. I. Katz\altaffilmark{}}
\affil{Department of Physics and McDonnell Center for the Space Sciences}
\affil{Washington University, St. Louis, Mo. 63130}
\email{katz@wuphys.wustl.edu}
\begin{abstract}
Chirped radio pulses known as perytons are consistent with plasma
propagation, but an astronomical origin has been excluded.  This paper
considers several physical processes as possible explanations.  TE and TM
waveguide modes have dispersion relations formally similar to those of
plasmas, and emission at the plasma frequency of a plasma recombining by
two-particle processes also produces a chirp $\delta t \propto \omega^{-2}$,
but these hypotheses fail quantitatively.  The chirp might be approximately
fitted to three-body recombination, cyclotron radiation in exponentially
decaying fields or to the resonance of a distributed LC circuit whose
component conductors move.  Constraints are placed on these possible
processes, but no specific source has been identified.
\end{abstract}
\maketitle
\section{Introduction}
Perytons \citep{BS11,K12,BNM12,SBM14,KONZJ14} are short (tens of ms) chirped
radio pulses first observed in the Parkes Multibeam Pulsar Survey.  The most
remarkable property of perytons is that their chirps have a very similar
functional dependence of frequency on time to those of radio pulsars $\delta
t \propto \omega^{-2}$ and apparently extragalactic Fast Radio Bursts
\citep{L07,T13}.  The dispersion measures of perytons can be regarded as
measures of their chirp rates independent of their cause, and are almost all
between 350 and 400 pc-cm$^{-3}$.  These values, but not their
concentration in a narrow range, are characteristic of propagation from
cosmological distances \citep{T13}.  A simple calculation shows that they
are inconsistent with emission from sources of stellar or lesser dimensions
because of free-free absorption \citep{S62} at the high densities that would
be implied if the dispersion were produced along such short path lengths.  A
variety of arguments indicate that perytons are not astronomical events, but
they remain of interest to astronomers as a source of interference and of
possible misidentification as astronomical phenomena.

The purpose of this paper is to consider possible telluric (for example, in
lightning or meteor trails) or anthropogenic origins of the chirped signals
of perytons.  Propagation in a waveguide (except for TEM modes) and emission
by a plasma recombining by two-body processes produce chirps $\delta t
\propto \omega^{-2}$ mimicking that of plasma propagation.  Three-body
recombination or cyclotron radiation in the magnetic field of an impulsively
excited LR circuit lead to different functional forms but may be fitted to
plasma dispersion chirps over limited ranges of frequency.  The resonant
frequencies of LC circuits whose values of inductance $L$ or capacitance $C$
change with time are chirped, as are the resonant frequencies of conducting
structures whose geometry changes, but special assumptions must be made to
reproduce the observed chirps of perytons.  In any of these models the
instantaneous emission bandwidth (typically $\Delta \nu/\nu \approx 0.03$) 
is a measure of the homogeneity of the emission region.  In this paper I
consider each of these physical processes as a possible source of perytons.
\section{Propagation}
\subsection{Cold Plasma}
The dispersion relation for electromagnetic waves in cold plasma is
\begin{equation}
\omega^2 = \omega_p^2 + c^2 k^2,
\label{plasmadisp}
\end{equation}
where the plasma frequency 
\begin{equation}
\omega_p \equiv \sqrt{4 \pi n e^2/m},
\label{plasmafreq}
\end{equation}
and $n$ and $m$ are the number density and mass of the species (usually
electrons, but possibly other charged particles) dominating the plasma
dispersion, $\omega$ is the wave angular frequency, $k$ is the wave vector,
$c$ the speed of light and $e$ the unit of charge.  The propagation time is
\begin{equation}
t = \int\,{d\ell \over v_g}
\end{equation}
where the group velocity of the wave $v_g = d\omega/dk$ and the integral is
over path elements $d\ell$.  Then
\begin{equation}
t = \int\,{d\ell \over c}\left(1 + {1 \over 2}\left({\omega_p \over \omega}
\right)^2 + {3 \over 8}\left({\omega_p \over \omega}\right)^4 + \cdots
\right).
\end{equation}

It is usually (for example, in intergalactic and interstellar media, even in
dense clouds) sufficient to take the first two terms.  The third term may be
significant in denser plasma, such as a stellar corona, but at such high
density inverse bremsstrahlung absorption \citep{S62} precludes the
observation of any emitted radiation.  As is frequently the case in such
expansions, when higher terms are significant the expansion is so slowly
convergent that the exact expression must be used.

The first term is independent of frequency and does not contribute to 
dispersion.  Then
\begin{equation}
\delta t(\omega) \equiv t(\omega) - t(\infty) \approx {1 \over 2} {4 \pi e^2
\over m_e \omega^2 c} {\rm DM},
\label{plasmadelay}
\end{equation}
where the dispersion measure ${\rm DM} \equiv \int d\ell\, n$ and the
subscript $e$ refers to electrons.  The chirp of most perytons is well, but
not exactly, fitted by (\ref{plasmadelay}) with DM in the narrow range
350--400 pc/cm$^3$, corresponding to a delay at wave frequency $\nu \equiv
\omega /2\pi = 1400\,$MHz of $\delta t_{1400} = 0.75$--0.86 s.
\subsection{Waveguides}
A chirp with the frequency dependence of (\ref{plasmadelay}) may be
arise from other causes.  The dispersion relation of waves in a simply
connected waveguide is
\begin{equation}
\omega^2 = \omega_0^2 + c^2 k^2,
\label{waveguidedisp}
\end{equation}
where $\omega_0$ is the cutoff angular frequency of the mode of propagation
considered.  Eq.~\ref{waveguidedisp} leads to a relation of the same form as
Eq.~\ref{plasmadelay}, with indistinguishable consequences:
\begin{equation}
\delta t(\omega) \approx {1 \over 2} {\ell \over c} {\omega_0^2 \over
\omega^2}.
\label{waveguidedelay}
\end{equation}
The equivalence between $\omega_0$, the waveguide length $\ell$ and the
dispersion measure is
\begin{equation}
\ell \omega_0^2 \Longleftrightarrow {4 \pi e^2 \over m_e}\,{\rm DM} = 3.2
\times 10^9 {\rm {cm^3 \over s^2}\,DM}.
\end{equation}

Unfortunately, propagation in a waveguide cannot explain chirped perytons
because the required length would be
\begin{equation}
\ell \approx 2 c \delta t_{1400} {\omega^2 \over \omega_0^2}
\gtrsim 10^{11}\,{\rm cm}.
\end{equation}
There are no such waveguides, and if they did exist they would be orders of
magnitude longer than their attenuation lengths that are typically O$(10^4 c
/\omega) \approx 3 \times 10^4\,$cm \citep{J99}.
\section{Plasma recombination}
An alternative explanation for chirped perytons may be sought in plasma
recombination in a spark (natural lightning or anthropogenic, perhaps in
electrical equipment) or discharge, a meteor entry trail, or some other
natural or artificial phenomenon.  A variety of processes, including
beam-plasma instability, excite electrostatic plasma waves with frequencies
close to $\omega_p$.  At density gradients these waves couple to propagating
electromagnetic waves.  These processes are familiar in laboratory plasmas.
If the plasma density declines with time, or if the excitation moves from
regions of greater to lesser plasma density \citep{BS11}, so will
$\omega_p$, producing a chirp that could under the right conditions be
indistinguishable from plasma dispersion.  The narrow instantaneous
frequency width of the emission would require remarkable homogeneity (about
5\%) in the density of emitting plasma.
\subsection{Two-body recombination}
Recombination reduces the charged particle density of a plasma.  If
recombination is the result of two-body reactions between positively and
negatively charged species (for example, electrons and positive atomic or
molecular ions):
\begin{equation}
{dn \over dt} = - \alpha_2 n^2,
\label{kinetics}
\end{equation}
where $\alpha_2$ is the recombination coefficient and we have assumed there
are two singly charged species, each with the number density $n$ but
opposite charges, as required by electrostatic neutrality, and that there is
no further ionization.

The trivial solution to Eq.~\ref{kinetics} is 
\begin{equation}
n(t) = {1 \over \alpha_2 t},
\label{plasmadensity}
\end{equation}
where $t$ (corresponding to $\delta t$ in Eqs.~\ref{plasmadelay},
\ref{waveguidedelay}) is the time elapsed since the nominal creation of the
plasma at infinite density.  Equating the observed radiation frequency
$\omega$ to $\omega_p$, using Eqs.~\ref{plasmafreq} and \ref{plasmadensity},
we find
\begin{equation}
\delta t = {4 \pi e^2 \over m \alpha_2 \omega^2}.
\label{kineticsdelay}
\end{equation}
This has the same form as the dispersion delays Eqs.~\ref{plasmadelay} and
\ref{waveguidedelay} with the equivalence 
\begin{equation}
\alpha_2 \Longleftrightarrow 2 {m_e \over m} {c \over {\rm DM}},
\end{equation}
and appears to be an alternative explanation of the chirp.  It even offers a
plausible explanation of deviations observed in perytons from the exact
$\delta t \propto \omega^{-2}$ because $\alpha_2$ depends on temperature,
which is likely to vary in atmospheric or electronic systems, and the
kinetics may be complicated by the presence of other species; many other
possible explanations (or excuses) exist.

Unfortunately, the numbers don't work out.  We distinguish two cases:
\begin{itemize}
\item For an electron-ion plasma $m = m_e$ and the required $\alpha_2
\approx 5 \times 10^{-11}\,$cm$^3$/s.  Because the electrons are unbound to
the ions that may capture them, two-body recombination requires the emission
of a photon and is slow.  At temperatures O$(1\,$eV), as found in sparks and
lightning, the radiative recombination rate $\alpha_2 \approx 3 \times
10^{-13}\,T_e^{-1/2}$, where the electron temperature $T_e$ is in units of
eV \citep{NRL}.  
\item In an ionic plasma (such as may result in weakly ionized air
if electrons are captured by neutral molecules to make negative molecular
ions) recombination may occur in non-radiative collisions because the excess
energy is taken up by the reaction products' kinetic energy.  Then $m$ is
properly the reduced mass of the two ions ($\approx 14$ a.m.u.) and the
required $\alpha_2 \approx 2 \times 10^{-15}$ cm$^3$/s.  A plasma of such
massive ions must be very dense to have the inferred plasma frequency, and
hence $\alpha_2$ must be small to keep the recombination rate in the range
inferred from the observed chirp.  Ionic collisional neutralization (charge
exchange) rates are not small because the cross-sections are O($5 \times
10^{-15}$ cm$^2$) \citep{F96}; $\alpha_2 \approx 10^{-9}\,T^{1/2}$ cm$^3$/s,
with $T$ in eV.  Similar results apply to atomic ions.
\end{itemize}
In neither case is it possible to obtain a recombination coefficient
consistent with observed peryton chirps.
\subsection{Three-body recombination}
\label{3B}
At higher electron densities radiative recombination is insignificant
compared to three-body recombination.  Eq.~\ref{kinetics} is replaced by
\begin{equation}
{dn \over dt} = - \alpha_3 n^3,
\end{equation}
where $\alpha_3$ is the three-body recombination coefficient, if the third
body is a second electron and the recombining ion density equals the
electron density (there are no other charged particles)\footnote{An equation
analogous to Eq.~\ref{kinetics}, with $\alpha_2$ replaced by
$\alpha_{3B}n_B$, is obtained if the third body is a bystander, such as the
dominant neutral species, whose density $n_B$ is nearly constant.  This may
apply to weakly ionized gases.}.  Then $\delta t \propto n^{-2} \propto
\omega^{-4}$ rather than $\delta t \propto n^{-1} \propto \omega^{-2}$
(Eq.~\ref{plasmadelay}).

Over a limited frequency range a fit to $\delta t = C_2 \omega^{-2}$ may not
be distinguishable from a fit to $\delta t = C_4 \omega^{-4}$ with $C_4 =
C_2 {\bar \omega}^2/2$, equating the chirp rates at $\bar \omega$, the
midpoint of the range.  The required three-body recombination coefficient
\begin{equation}
\alpha_3 = {32 \pi e^2 c \over m_e {\bar \omega}^2 {\rm DM}} \approx 
8 \times 10^{-21}\ {\rm cm^6/s}.
\end{equation}
This may be compared to the estimate $\alpha_3 \approx 9 \times 10^{-27}\,
T_e^{-4.5}$, where $T_e$ is in eV \citep{NRL}.  These rates might be
reconciled for $T_e \approx 0.05\,$eV (500$^{\,\circ}$K, a plausible value
for a weakly ionized gas), but assumption-dependent simulations of complex
reaction networks would be required for quantitative conclusions, as well as
a quantitative test of how well peryton chirps may be fit by $\delta t
\propto \omega^{-4}$ rather than $\delta t \propto \omega^{-2}$.  In
addition, the strong temperature dependence implies that unavoidable
temperature changes in the course of recombination will make the actual
$\delta t(\omega)$ deviate from a simple power law.
\section{Cyclotron radiation}
\label{cyclotron}
The frequency of cyclotron radiation by an electron in a magnetic induction
$B$ is
\begin{equation}
\nu = {e B \over 2 \pi m_e c}.
\end{equation}
Observed peryton frequencies $\nu \approx 1400\,$MHz correspond to $B
\approx 500$ gauss, a field readily achievable with permanent or
electromagnets.  The observed chirp $\delta t \propto \omega^{-2}$ requires
$B \propto (\delta t)^{-1/2}$.  If the field is produced by a linear
ferromagnetic material, the current magnetizing it must also vary $\propto
(\delta t)^{-1/2}$.  This does not have an obvious origin.  Currents in
simple LR circuits decay exponentially with a time constant $\tau = L/R$, 
where $L$ is the inductance and $R$ the resistance.  Inclusion of
ferromagnetic material such as an electromagnet in a circuit implies a large
$L$ and a slow decay of field and current.  Ferromagnetic materials are
generally very nonlinear, and a simple functional form for the decay of the
magnetic field would not be expected.

Just as for three body recombination, any smooth decay over a limited
frequency range can be fitted to a $\delta t \propto \omega^{-2}$ form.
Fitting an exponential decay to a slope $dt/d\omega$ described by the
dispersion measure DM implies
\begin{equation}
\tau = {4 \pi e^2 \over m_e \omega^2 c}{\rm DM} \approx 1.6\,{\rm s}.
\end{equation}
This is consistent with plausible values of $L$ and $R$, but there is no
reason to expect such circuits to have this particular value of $\tau$.

The narrowness ($\Delta \nu/\nu \approx 0.03$) of the instantaneous spectrum
constrains the field homogeneity in an emission region.  Homogeneity may not
be as serious an issue as for plasma recombination models because devices
may be engineered to have homogeneous magnetic fields in the plasma region,
while within a spark the plasma frequency varies continuously from zero 
outside to its maximum.

We conclude that although the magnetic fields necessary for cyclotron
radiation at the observed peryton frequencies exist in terrestrial magnets,
the observed chirp is not a natural product of electromagnet circuits.
\section{LC Circuits}
The oscillation frequency of an LC circuit
\begin{equation}
\omega = {1 \over \sqrt{LC}}.
\end{equation}
If the product $LC \propto \delta t$ then $\omega$ decreases from a nominal
infinite value at $\delta t = 0$ according to a relation
\begin{equation}
\delta t = \left(\omega^2 {d(LC) \over dt}\right)^{-1}
\end{equation}
analogous to Eq.~\ref{plasmadelay} with
\begin{equation}
{d(LC) \over dt} \Longleftrightarrow {m_e c \over 2 \pi e^2 {\rm DM}} = 19
{\rm {s \over cm^2}\,{1 \over DM}} \approx 1.6 \times 10^{-20}\,{\rm s}.
\label{LCderiv}
\end{equation}

A circuit with the required resonant $\omega = {\rm O}(10^{10}$/s) has
dimensions $\lesssim c/\omega = 3\,$cm, no lumped impedances, and
distributed $L \approx 10\,$nH and $C \approx 1\,$pF.  If its geometry
changes so will $LC$.  Although the functional form $\delta t \propto
\omega^{-2}$ requires a constant $d(LC)/dt$, just as in Sections \ref{3B}
and \ref{cyclotron}, it may be possible to fit the observed peryton
$d\omega/dt$ over a limited frequency range even if $d(LC)/dt$ varies,
provided it approximates Eq.~\ref{LCderiv} during the limited period of
observation.

The required circuit must change its geometry, and therefore its resonant
frequency, substantially in a few tenths of a second.  In addition, some
mechanism, such as a spark, must excite its resonant electromagnetic
oscillations.  For example, we might consider a bimetallic strip in a
thermostat that changes its shape as the temperature relaxes, or a 
mechanical relay, and a spark that is struck as the gap opens.  Varying
electromagnetic resonance is consistent with perytons, provided circuits
with the required resonant frequency and chirp exist, but none are yet
known.
\section{Discussion}
We have considered several processes that might explain the most remarkable
property of perytons, their chirp that resembles that of plasma dispersion
at extragalactic distances.  Each of these proposals encounters difficulty,
qualitative (difficulty in explaining the functional form of the chirp) or
quantitative (difficulty in explaining the magnitude of the chirp).  The LC
circuit may be the most promising, if a suitable mechanically relaxing
configuration exists.

No sources of perytons have been identified.  Their narrow range of
dispersion measures and observed timing patterns (including preferred times
of day and intervals between events \citep{BS11,K12}) and the absence of an
association with any known natural telluric phenomenon point to an
anthropogenic origin in a single class of electronic equipment, perhaps in
an unexpected or abnormal state.  Although perytons are local to the
observers, they have been observed in both Australia and Switzerland,
indicating that their sources are widely distributed.  However, no perytons
were observed at the Allen Telescope Array in California when $\sim 10$
would have been expected based on their event rate observed at Parkes
\citep{S12}, indicating that their sources are not ubiquitous.

The perytons observed at Parkes were detected in all 13 beams
of the multibeam receiver.  Possible explanations include: 1) It was so
close ($\lesssim 10\,$km) to the telescope that it was out of focus
\citep{KONZJ14}; 2) The emission was distributed over a two-dimensional
region on the sky at least a few degrees wide and long; 3) It was observed
in the far sidelobes of all the beams \citep{BS11}; 4) The signal entered by
``back door'' direct coupling to the circuits rather than by amplification
of electromagnetic energy collected and focused by the telescope.

Speculative possibilities include devices, such as cathode ray tubes,
amplifiers and power supplies, associated with radio telescopes, components
of the power grid such as circuit breakers and voltage regulators, and 
consumer devices such as microwave ovens and fluorescent and vapor lamps.
The time scale described by their chirp may be related to that of 
electromechanical switches.  It may be fruitful to search for anthropogenic
peryton-like events with simple dipole antennas and receivers in a variety
of cultural environments, not limited to radio telescopes.  If they are the
result of ``back door'' coupling, the giant focusing dishes of radio
telescopes would not be necessary for their detection.
\acknowledgments
I thank P. Horowitz and S. R. Kulkarni for useful discussions.


\begin{thebibliography}{9}
\bibitem[\protect\citeauthoryear{Bagchi, Nieves \& McLaughlin}{2012}]{BNM12}
Bagchi, M., Nieves, A.~C. \& McLaughlin, M. 2012 \mnras\ 425, 2501.
\bibitem[\protect\citeauthoryear{Burke-Spolaor, {\it et al.\/}}{2011}]{BS11}
Burke-Spolaor, S., Bailes, M., Ekers, R., Marquart, J.-P. \& Crawford, 
F.~III 2011 \apj\ 727, 18
\bibitem[\protect\citeauthoryear{Flannery}{1996}]{F96}Flannery, M.~R. {\it
Atomic, Molecular \& Optical Physics Handbook\/} ed. G.~W.~F.~Drake (AIP
Press, New York 1996).
\bibitem[\protect\citeauthoryear{Kocz, {\it et al.\/}}{2012}]{K12}Kocz, J.,
Bailes, M., Barnes, D., Burke-Spolaor, S. \& Levin. L. 2012 \mnras\ 420,
271.
\bibitem[\protect\citeauthoryear{Jackson}{1999}]{J99}Jackson, J.~D. {\it
Classical Electrodynamics\/} 3rd ed. (Wiley, New York 1999).
\bibitem[\protect\citeauthoryear{Kulkarni, {\it et al.\/}}{2014}]{KONZJ14}
Kulkarni, S.~R., Ofek, E.~O., Neill, J.~D., Zheng, Z. \& Juric, M. 2014 
arXiv:1402.4766.
\bibitem[\protect\citeauthoryear{Lorimer, {\it et al.\/}}{2007}]{L07}
Lorimer, D.~R., Bailes, M., McLaughlin, M.~A., Narkevic, D.~J., \& Crawford,
F. 2007 Science 318, 777.
\bibitem[\protect\citeauthoryear{NRL}{2007}]{NRL}{\it NRL Plasma Formulary}
2007 Naval Research Laboratory, Washington, D.~C.
\bibitem[\protect\citeauthoryear{Saint-Hilaire, Benz \& Monstein}{2014}]
{SBM14}Saint-Hilaire, P., Benz, A.~O. \& Monstein, C. 2014 arXiv:1402.0664.
\bibitem[\protect\citeauthoryear{Siemion, {\it et al.\/}}{2012}]{S12}
Siemion, A.~P.~V., {\it et al.\/} 2012 \apj\ 744, 109.
\bibitem[\protect\citeauthoryear{Spitzer}{1962}]{S62}Spitzer, L. Jr. {\it 
Physics of Fully Ionized Gases\/} (Interscience, New York 1962).
\bibitem[\protect\citeauthoryear{Thornton, {\it et al.\/}}{2013}]{T13}
Thornton, D., {\it et al.\/} 2013 Science 341, 53.
\end{thebibliography}
\end{document}